\documentclass[12pt,preprint]{elsarticle}
\usepackage{amsmath}
\usepackage{amssymb}

\usepackage{color}

\newtheorem{theorem}{Theorem}[section]

\newtheorem{remark}[theorem]{Remark}\numberwithin{equation}{section}

\begin{document}

\begin{frontmatter}

\title{Group classification of the two-dimensional Green-Naghdi equations with a time dependent bottom topography}

\author[SUT]{S.V.~Meleshko \corref{cor1}}

\cortext[cor1]{Corresponding author}

\ead{sergey@math.sut.ac.th}

\address[SUT]{School of Mathematics, Institute of Science, Suranaree University of Technology, Nakhon Ratchasima, 30000, Thailand}

\begin{abstract}
The two-dimensional Green-Naghdi equations with uneven bottom topography are studied in this paper. The function defining the bottom topography can be dependent on time. Group classification of these
equations with respect to the function describing the topography of
the bottom is performed in the paper. The algebraic approach used for the analysis of the classifying equations.
\end{abstract}
\end{frontmatter}

\section{Introduction}

Mathematical modeling of physical phenomena is one of the main streams
in continuum mechanics. Phenomena such as hydraulic currents, coastal
currents, currents in rivers and lakes, currents in water intakes,
tsunami simulation, breakout wave propagation, distribution of heavy
gases and scales of atmospheric movements used in weather forecasting
require mathematical consideration.

Motion of an ideal fluid flow under the force of gravity can be modeled
by means of the Euler equations. However, the full Euler equations,
even under the assumption of incompressibility, barotropy and absence
of rotation, are still rather complicated for describing waves on
a surface. One of these difficulties is that the free surface is a
part of the solution. This difficulty has motivated scientists to
derive simpler equations. For this reason development of approximate
models and their analysis by analytical and numerical methods is an
actual problem.

The need to reduce the original equations to simpler equations led
to the construction of asymptotic expansion models with respect to
a small parameter determined by the ratio of the depth of the fluid
to the characteristic linear size. One class of such equations is
the class of shallow water equations.

The shallow water equations describe the motion of an incompressible
fluid in the gravitational field if the depth of the fluid layer is
sufficiently small. They are widely used in the description of processes
in the atmosphere, water basins, modeling of tidal oscillations, tsunami
waves and gravitational waves~(e.g., see the classical books such
as~\cite{bk:Whitham[1974],bk:Pedlosky,bk:Ovsiannikov_C,bk:Salmon[1998]}
and \cite{bk:Vallis2006,bk:PetrosyanBook2010}).

There are many approaches for
deriving shallow water models, a review of which can be found in
\cite{bk:BonnetoBarthelemy_et[2011],bk:KhakimzyanovDutykhFedotovaMitsotakis}.
The classical approach of deriving the shallow water equations consists
of approximating the Euler equations for irrotational flows. The hierarchy
of the shallow water approximations is considered with respect to
the shallowness parameter $\delta=h_{0}/L$, where $h_{0}$ is the
mean depth of the fluid, and $L$ is the typical length scale of the
wave \cite{bk:Matsuno[2015]}\footnote{See also the references therein.}.
In particular, the Green-Naghdi equations, derived for describing
the two-dimensional fluid flow over an uneven bottom, are accurate
up to the dispersive terms of order $\delta^{2}$. The Green-Naghdi
system of equations is the generalization of the equations derived
first by Serre \cite{bk:Serre[1953]} and later by Su and Garden
\cite{bk:SuGardner[1969]} to describe the one-dimensional propagation
of fully nonlinear and weakly dispersive surface gravity waves over
a flat bottom topography.

In the classical Green-Naghdi equations, there are no equivalence transformations defined by Galilean invariance.
To overcome this obstacle, the Green-Naghdi equations with
bottom topography depending on time were derived in \cite{bk:KaptsovMeleshkoSamatova2021}. For deriving these equations the authors used Matsuno's approach \cite{bk:Matsuno[2015]}. It should be noted that the equations derived in this manner coincide
with the equations obtained by a different approach\footnote{Private communication with S.L.Gavrilyuk.}.
The dependence on time can take into account the bottom motion during, for example, an under water
earthquake or moving object located on the bottom. Some experimental
and theoretical results for two specific time deformations of the
bottom are presented in \cite{bk:Hammack1973}. The authors of \cite{bk:DutykhDias2007b}
considered motion of the bottom defined by the formula $\zeta(x,y)T(t)$.
Using the Fourier transform with respect to $(x,y)$ and the Laplace
transform in $t$ of the linearized problem, analytical formulas for
the free-surface elevation was derived there.

It is well-known that symmetries of a mathematical model are intrinsic
properties inherited from physical phenomena. One of the tools for
studying symmetries is the Lie group analysis method \cite{bk:Ovsiannikov1978,bk:Olver[1986]},
which is a basic method for constructing exact solutions of ordinary
and partial differential equations. Even in the case of the one-dimensional
shallow water equations for flat bottom, one encounters certain difficulties
to obtain nontrivial exact solutions. Symmetries of differential equations
yield a guarantied source of group-invariant solutions. Applications
of Lie groups to differential equations is the subject of many books
and review articles~\cite{bk:Ovsiannikov1978,bk:Olver[1986],bk:Ibragimov[1983],bk:BlumanKumei1989,
bk:HandbookLie,bk:Gaeta1994}.

Applications of the group analysis method to the Green-Naghdi equations
with a horizontal bottom topography in Eulerian and Lagrangian coordinates
were studied in \cite{bk:BagderinaChupakhin[2005],bk:SiriwatKaewmaneeMeleshko2016}.
In \cite{bk:SiriwatKaewmaneeMeleshko2016,bk:DorodnitsynKaptsovMeleshko2021},
Noether's theorem is applied for finding conservation laws of the
one-dimensional classical Green-Naghdi equations with horizontal and
uneven bottom topography, respectively.

Group properties of the one-dimensional Green-Naghdi equations with uneven bottom depending on time were studied in \cite{bk:KaptsovMeleshkoSamatova2021}.
One of the main focuses of \cite{bk:KaptsovMeleshkoSamatova2021} was the representation of one-dimensional equations in a variational form.
For this purpose, the equations were
rewritten in mass Lagrangian coordinates. The variational form allowed using Noether’s theorem for constructing conservation laws.
Because the first step of application
of Noether’s theorem is the group analysis of the Euler-Lagrange equation, the complete group classification of the studied equations was performed.

The present paper is devoted to the group classification of the two-dimensional Green-Naghdi equations with respect to the bottom topography depending on time. The equations are considered in Eulerian coordinates. For the group classification we use the algebraic approach applied earlier to different types of systems (see, e.g., \cite{bk:GrigorievMeleshkoSuriyawichitseranee2012,bk:Chirkunov[2012],bk:Kasatkin[2012],bk:MkhizeMoyoMeleshko[2014],bk:OpanasenkoBoykoPopovych_2020} and references therein).
The algebraic approach takes the algebraic properties of an admitted
Lie algebra into account and allows one a significant simplification
of the group classification. It was applied in the case if the admitted generators belong to subalgebras of some Lie algebra.

\bigskip

The paper is organized as follows.
In Section~\ref{sec:diffeqns}, the Green-Naghdi equations with uneven bottom depending on time are given. Equivalence transformations are presented in Section~\ref{sec:equivalence}.
Section~\ref{sec:Group_classification} is devoted to the group classification of the sudied equations.
The obtained results are discussed in the Conclusions.

\section{Studied equations}
\label{sec:diffeqns}

In the dimensional form, the two-dimensional Green-Naghdi equations
with a bottom topography depending on time have the form
\begin{equation}
\begin{array}{c}
\frac{\partial h}{\partial t}+\nabla(h\boldsymbol{u})=0,\\[2ex]
\frac{d}{dt}\boldsymbol{u}+g\nabla(h+H)=\frac{1}{h}\left(\nabla A+B\nabla H\right),
\end{array}\label{eq:SGN}
\end{equation}
where
\begin{equation}
A=h^{2}\frac{d}{dt}\left(\frac{h}{3}\,div(\boldsymbol{u})-\frac{d}{dt}H\right),\ \ B=h\frac{d}{dt}\left(\frac{h}{2}\,div(\boldsymbol{u})-\frac{d}{dt}H\right),\label{eq:original_bottom}
\end{equation}
and
\[
\frac{d}{dt}=\frac{\partial}{\partial t}+(\boldsymbol{u}\cdot\nabla).
\]
Here $h$ is the total depth of the fluid, $H$ is the height of the
fluid column between the bottom and the undisturbed level of the fluid
(see Figure \ref{fig:vars}). The function $H(x,y,t)$ describes the
variable bottom topography, e.g. the ocean bathymetry or a moving
object located on the bottom. This function is assumed to be known.
The first equation of the latter system is the conservation law of
mass of the fluid column over which the averaging occurs, and the
second equation is the Eulerian form of the Newton's second law, with
contributions to the fluid particle acceleration due to gravity and
due to the surface and bottom boundary conditions.

Notice that for $H_{t}=0$ these equations coincide with the Green-Naghdi
equations \cite{Bazdenkov_C_1987,bk:LiapidevskiiGavrilova2008,bk:Matsuno[2016]}\footnote{It should be also noted that even $H_{t}=0$ in \cite{bk:LiapidevskiiGavrilova2008},
their Green-Naghdi equations are written in the form (\ref{eq:SGN})--(\ref{eq:original_bottom}).}.

\section{Equivalence transformations}
\label{sec:equivalence}

The class of equations (\ref{eq:SGN}) is parameterized by the arbitrary
elements $H(x,y,t)$. Equivalence transformations of this class preserve
the structure of its equations, but are allowed to change the arbitrary
elements. The first step of the group classification of the class
of equations of the form (\ref{eq:SGN}) is to describe the equivalence
among equations of this class to which the group classification is
carried out.

Generators of one-parameter groups of equivalence transformations
are assumed to be in the form \cite{bk:Ovsiannikov1978,bk:Meleshko[2005]}
\[
X^{{\rm e}}=\xi^{t}\partial_{t}+\xi^{x}\partial_{x}+\xi^{y}\partial_{y}+\eta^{h}\partial_{h}+\eta^{u}\partial_{u}+\eta^{v}\partial_{v}+\zeta^{H}\partial_{H},
\]
where all coefficients of the generator depend on $(t,x,y,h,u,v,H)$.

The class of differential equations (\ref{eq:SGN}) is
defined by auxiliary equations for the arbitrary elements $H$, which
are given by
\[
H_{h}=0,\ \ H_{u}=0,\ \ H_{v}=0.
\]

For finding equivalence transformations the infinitesimal criterion
\cite{bk:Ovsiannikov1978} is used. For this purpose the determining
equations for the components of generators of one-parameter groups
of equivalence transformations were derived. The solution of these
determining equations gives the general form of elements of the equivalence
group of the class (\ref{eq:SGN}). Because of the cumbersome calculations
we extend the equivalence transformations, found in \cite{bk:KaptsovMeleshkoSamatova2021}
for one-dimensional case, to the two-dimensional case
\begin{equation}\label{eq:equiv_gen}
\begin{array}{c}
X_{1}^{e}=\partial_{x},\ X_{2}^{e}=\partial_{y},\ X_{3}^{e}=t\partial_{x}+\partial_{u},\ X_{4}^{e}=t\partial_{y}+\partial_{u},\
\\[2ex]
X_{5}^{e}=-y\partial_{x}+x\partial_{y}-v\partial_{u}+u\partial_{v},\ X_{6}^{e}=\partial_{t},
\\[2ex]
X_{7}^{e}=t\partial_{t}+2x\partial_{x}+2y\partial_{y}+u\partial_{u}+v\partial_{v}+2h\partial_{h}
\\[2ex]
X_{8}^{e}=t\partial_{t}-2g\partial_{g},\,\,\,X_{9}^{e}=h\partial_{h}+g\partial_{g}+H\partial_{H},\,\,\,X_{10}^{e}=\partial_{H},\,\,\,X_{11}^{e}=t\partial_{H}.
\end{array}
\end{equation}

The corresponding transformations changing $H$ are
\[
X_{10}^{e}:\,\,\,\,\tilde{H}=H+a,
\]
\[
X_{11}^{e}:\,\,\,\,\,\tilde{H}=H+at,
\]
where $a$ is the group parameter. Hence, because of the transformations
related to $X_{10}^{e}$ and $X_{11}^{e}$, if $H(x,y,t)=G(x,y,t)+w(t)$
and $w^{\prime\prime}=0$, then one can assume that $w=0$.

There are also two obvious involutions which correspond to the change
of the variables:
\begin{equation}
\begin{array}{c}
E_{1}:\,\,\,\,\tilde{x}=-x,\,\,\,\tilde{y}=-y,\,\,\,\tilde{u}=-u,\,\,\,\tilde{v}=-v,\\
E_{2}:\,\,\,\,\,\,\,\tilde{t}=-t,\,\,\,\tilde{u}=-u,\,\,\,\tilde{v}=-v,
\end{array}\label{eq:dec07.21.1}
\end{equation}
where only changeable variables are presented.

\begin{remark} The transformation corresponding to the generator
$X_{3}^{e}$ and $X_{4}^{e}$ are the Galilean transformations. Notice
that the Green-Naghdi equations, when $H$ does not depend on time
$t$, do not admit the Galilean transformations as equivalence transformations.
The Galilean invariance principle states that all mechanical laws
are the same in any inertial frame of reference. This property is
of fundamental importance for any mathematical model.

\end{remark}

\section{Group classification}
\label{sec:Group_classification}

Group classification is carried out up to equivalence transformations.
The group classification problem consists of finding all Lie algebras
admitted by equation (\ref{eq:SGN}). A part of these Lie algebras,
called the kernel of admitted Lie algebras, is admitted for all arbitrary
element $H$. Another part depends on the specification of the arbitrary
elements. This part contains nonequivalent extensions of the kernel
of admitted Lie algebras.

The admitted Lie algebra consists of the generators
\[
X^{{\rm e}}=\xi^{t}\partial_{t}+\xi^{x}\partial_{x}+\xi^{y}\partial_{y}+\eta^{h}\partial_{h}+\eta^{u}\partial_{u}+\eta^{v}\partial_{v}+\zeta^{H}\partial_{H},
\]
where all coefficients of the generator depend on $(t,x,y,h,u,v,H)$
and satisfy the determining equations \cite{bk:Ovsiannikov1978}.

Calculations, performed in the symbolic manipulation system Reduce
\cite{bk:Hearn}, show that the classifying equations are
\begin{equation}
\frac{\partial S}{\partial x}=0,\,\,\,\frac{\partial S}{\partial y}=0,\,\,\,\frac{\partial^{2}S}{\partial t^{2}}=0,
\label{eq:dec06.21.1}
\end{equation}
where $x_{i}$, $(i=1,2,...,7)$ are constant,
\[
\begin{array}{c}
S=(x_{1}+x_{3}t)H_{x}+(x_{2}+x_{4}t)H_{y}+x_{5}(H_{y}x-H_{x}y)\\
+x_{6}H_{t}+x_{7}(H_{t}t+2H_{y}y+2H_{x}x-2H),
\end{array}
\]
and
\[
X=\sum_{i=1}^{7}x_{i}X_{i},
\]
with the generators
\[
\begin{array}{c}
X_{1}=\partial_{x},\ X_{2}=\partial_{y},\ X_{3}=t\partial_{x}+\partial_{u},\ X_{4}=t\partial_{y}+\partial_{u},\ \\
X_{5}=-y\partial_{x}+x\partial_{y}-v\partial_{u}+u\partial_{v},\ X_{6}=\partial_{t},\\
X_{7}=t\partial_{t}+2x\partial_{x}+2y\partial_{y}+u\partial_{u}+v\partial_{v}+2h\partial_{h}.
\end{array}
\]

There are several ways to analyze the classifying equations (\ref{eq:dec06.21.1}).
One of the methods has been applied for group classification of the
gas dynamics equations \cite{bk:Ovsiannikov1978}. Unfortunately,
implementation of this algorithm leads to cumbersome calculations.
An alternative method for analyzing the classifying equations consists
of an algebraic approach, taking the algebraic properties of an admitted
Lie algebra into account, thus allowing for a significant simplification
of the group classification. This algebraic approach for group classification
has been applied in \cite{bk:GrigorievMeleshkoSuriyawichitseranee2012,bk:Chirkunov[2012],bk:Kasatkin[2012],bk:MkhizeMoyoMeleshko[2014],bk:OpanasenkoBoykoPopovych_2020}\footnote{See also references therein.}.

As for the algebraic approach, one notes that the generators admitted
by equation (\ref{eq:SGN}) compose a Lie algebra which is a subalgebra
of the Lie algebra $L_{7}=\left\{ X_{1},X_{2},...,X_{7}\right\} $.
Another observation is that the automorphisms of the Lie algebra $L_{7}$
act similar to the equivalence transformations corresponding to the
generators $X_{i}^{e},\,\,\,(i=1,2,...,7)$. Hence, each of the Lie
algebras admitted by equation (\ref{eq:SGN}) belongs to one of the
classes of an optimal system of subalgebras of the Lie algebra $L_{7}$.
Thus, for the group classification of equation (\ref{eq:SGN}) it
is sufficient to construct an optimal system of subalgebras of the
Lie algebra $L_{7}$. Each representative of a class from the optimal
system of subalgebras provides a set of constants $x_{i},\,\,\,(i=1,2,...,7)$.
Using these constants, and substituting them into classifying equations
(\ref{eq:dec06.21.1}), one obtains an overdetermined system of equations
for the function $H(x,y,t)$. The general solution of the overdetermined
system of equations gives the group classification of equation (\ref{eq:SGN}).

\subsection{Optimal system of subalgebras of $L_{7}$}

For low-dimensional Lie algebras calculation of the optimal system
of subalgebras (also called the representative list of subalgebras)
is relatively easy. For high-dimensional Lie algebras the problem
becomes complicated because it requires extensive computations. The
difficulties can be facilitated by a two-step algorithm suggested
in~\cite{bk:Ovsiannikov[1993opt]}. This algorithm replaces the
problem of constructing the optimal system of high-dimensional subalgebras
by a similar problem for lower dimensional subalgebras.

Shortly the algorithm \cite{bk:Ovsiannikov[1993opt]} can be formulated
as follows. Let $L$ be a Lie algebra $L$ with the basis $\left\{ X_{1},X_{2},\ldots,X_{r}\right\} $.
Assume that the Lie algebra $L$ is decomposed as $L=I\oplus F$,
where $I$ is a proper ideal of the algebra $L$ and $F$ is a subalgebra.
Then the set of the inner automorphisms $A=\mbox{Int}\ L$ of the
Lie algebra $L$ is decomposed $A=A_{I}A_{F}$, where
\[
AI\subset I,\quad A_{F}F\subset F,\quad(A_{I}X)_{F}=X,\quad\forall X\in F.
\]
This means the following \cite{bk:Ovsiannikov[1993opt]}. Let $x\in L$
be decomposed as $x=x_{I}+x_{F}$, where $x_{I}\in I$, and $x_{F}\in F$.
Any automorphism $B\in A$ can be written as $B=B_{I}B_{F}$, where
$B_{I}\in A_{I},\;B_{F}\in A_{F}$. The automorphisms $B_{I}$ and
$B_{F}$ have the properties:
\[
\begin{array}{c}
B_{I}x_{F}=x_{F},\quad\forall x_{F}\in F,\quad\forall B_{I}\in A_{I},\\
B_{F}x_{I}\in I,\quad B_{F}x_{F}\in F,\quad\forall x_{I}\in I,\quad\forall x_{F}\in F,\quad\forall B_{F}\in A_{F}.
\end{array}
\]

At the first step, an optimal system of subalgebras $\Theta_{A_{F}}(F)=\{F_{0},F_{1},F_{2},...,F_{p},F_{p+1}\}$
of the algebra $F$ is formed. Here $F_{0}=\{0\}$, $F_{p+1}=\{F\}$
and the optimal system of the algebra $F$ is constructed with respect
to the automorphisms $A_{F}$. For each subalgebra $F_{j}$, $j=0,1,2,...,p+1$
one has to find its stabilizer $\mbox{St}(F_{j})\subset A$:
\[
\mbox{St}(F_{j})=\{B\in A\ |\ B(F_{j})=F_{j}\}.
\]
Note that $\mbox{St}(F_{p+1})=A$.

The second step consists of forming 
the optimal system of subalgebras $\Theta_{A}(L)$ of the algebra
$L$ as a collection of $\Theta_{\mbox{St}(F_{j})}(I\oplus F_{j})$,
$j=0,1,2,...,p+1$.

If the subalgebra $F$ can be decomposed, then the two-step algorithm
can be used for construction of $\Theta_{A_{F}}(F)$.

The structure of the Lie algebra is defined by its commutator table.
The commutator table of $L_{7}$ is
\begin{center}
\begin{tabular}{|c||c|c|c|c|c|c|c|}
\hline
 & $X_{1}$  & $X_{2}$  & $X_{3}$  & $X_{4}$  & $X_{5}$  & $X_{6}$  & $X_{7}$\\
\hline
$X_{1}$  & $0$  & $0$  & $0$  & $0$  & $X_{2}$  & $0$  & $2X_{1}$\\
\hline
$X_{2}$  & $0$  & $0$  & $0$  & $0$  & $-X_{1}$  & $0$  & $2X_{2}$\\
\hline
$X_{3}$  & $0$  & $0$  & $0$  & $0$  & $X_{4}$  & $-X_{1}$  & $0$\\
\hline
$X_{4}$  & $0$  & $0$  & $0$  & $0$  & $-X_{3}$  & $-X_{2}$  & $0$\\
\hline
$X_{5}$  & $-X_{2}$  & $X_{1}$  & $-X_{4}$  & $X_{3}$  & $0$  & $0$  & $0$\\
\hline
$X_{6}$  & $0$  & $0$  & $X_{1}$  & $X_{2}$  & $0$  & $0$  & $X_{6}$\\
\hline
$X_{7}$  & $-2X_{1}$  & $-2X_{2}$  & $0$  & $0$  & $0$  & $-X_{6}$  & $0$\\
\hline
\end{tabular}
\par\end{center}

Directly from the table one derives that the composition series of
ideals is
\[
O\subset\{X_{1},X_{2}\}\subset\{X_{1},X_{2},X_{3},X_{4}\}\subset
\]
\[
\{X_{1},X_{2},X_{3},X_{4},X_{5}\}\subset\{X_{1},X_{2},X_{3},X_{4},X_{5},X_{6}\}\subset L_{7},
\]
and inner automorphisms are:

\begin{table}
  \centering

\begin{tabular}{|c|c|c|c|c|c|}
\hline
$A$  & $Ap_{1}$  & $Ap_{2}$  & $Ax_{5}$  & $Ax_{6}$  & $Ax_{7}$ \\
\hline
$T$  & $p_{1}+2\alpha_{1}x_{7}+x_{5}(a_{1},-a_{2})$  & $p_{2}$  & $x_{5}$  & $x_{6}$  & $x_{7}$ \\
\hline
{\cal$\Gamma$}  & $p_{1}-x_{6}\alpha_{2}$  & $p_{2}+x_{5}(a_{3},-a_{4})$  & $x_{5}$  & $x_{6}$  & $x_{7}$ \\
\hline
$S$  & $Sp_{1}$  & $Sp_{2}$  & $x_{5}$  & $x_{6}$  & $x_{7}$ \\
\hline
$A_{6}$  & $p_{1}-a_{6}p_{2}$  & $p_{2}$  & $x_{5}$  & $x_{6}+a_{6}x_{7}$  & $x_{7}$ \\
\hline
$A_{7}$  & $\pm a_{7}^{2}p_{1}$  & $p_{2}$  & $x_{5}$  & $\pm a_{7}x_{6}$  & $x_{7}$ \\
\hline
${\it E_{1}}$  & $-p_{1}$  & $p_{2}$  & $x_{5}$  & $x_{6}$  & $x_{7}$ \\
\hline
${\it E_{2}}$  & $p_{1}$  & $-p_{2}$  & $x_{5}$  & $-x_{6}$  & $x_{7}$ \\
\hline
\end{tabular}
  \caption{Inner automorphisms of the Lie algebra $L_7$.}
  \label{tab:inner}
\end{table}

 where $a_{i}$, ($i=1,2,...,7$) are parameters of the automorphism
$A$,
\[
p_{1}=(x_{1},x_{2}),\ p_{2}=(x_{3},x_{4}),\ \alpha_{1}=(a_{1},a_{2}),\ \alpha_{2}=(a_{3},a_{4}),
\]
$S$ is the rotation matrix
\[
S=\left(\begin{array}{cc}
\cos a_{5} & \sin a_{5}\\
-\sin a_{5} & \cos a_{5}
\end{array}\right).
\]

For application of the two-step algorithm one notes that the Lie algebra
$L_{7}$ can be decomposed as follows
\[
L_{7}=\{\{\{\{\{X_{1},X_{2}\}\oplus\{X_{3},X_{4}\}\}\oplus\{X_{5}\}\}\oplus\{X_{6}\}\}\oplus\{X_{7}\}\}.
\]

\begin{remark}

 The Lie algebra $L_{7}$ coincides with the Lie algebra
admitted by the two-dimensional gas dynamics equations with the general
form of the state equation. The optimal system of subalgebras of the
Lie algebra $L_{7}$ was constructed by the author during the work
on the SUBMODELS program \citep{bk:Ovsiannikov[1994]}, leading by
L.V. Ovsiannikov. The two-step algorithm \cite{bk:Ovsiannikov[1993opt]}
was also proposed by L.V. Ovsiannikov in the framework of the SUBMODELS
program \citep{bk:Ovsiannikov[1994]}.
\end{remark}

The optimal system of subalgebras of the Lie algebra $L_{7}$ is presented
in Table \ref{Optimal_system}, where the subalgebra representatives
are denoted by a pair of numbers $(r,i)$: $r$ is the dimension and
$i$ is the serial number of a subalgebra of dimension $r$. The numbers
$r$ are given in front of each block containing sub-algebras of dimensions
$r$. The serial numbers $i$ are presented in the first column. The
bases of the subalgebras $(r,i)$ are written out in the second column.
\begin{table}[h!]
\centering

\begin{tabular}{|c|c||c|c|}
\hline
N  & Basis  & N  & Basis \\
\hline
\hline
\multicolumn{2}{|c||}{r=7} & \multicolumn{2}{c|}{r=3}\\
\hline
1  & $X_{1},X_{2},X_{3},X_{4},X_{5},X_{6},X_{7}$  & 11  & $X_{1},X_{2},X_{6}$ \\
\hline
\hline
\multicolumn{2}{|c||}{r=6} & \multicolumn{1}{c|}{12 } & $X_{1},X_{2},X_{3}+X_{6}$ \\
\hline
1  & $X_{1},X_{2},X_{3},X_{4},X_{5}+\alpha X_{7},X_{6}$  & 13  & $X_{3},X_{4},X_{7}$ \\
\hline
2  & $X_{1},X_{2},X_{3},X_{4},X_{5},X_{7}$  & 14  & $X_{1},X_{3}+\alpha X,X_{7}+\beta X_{4}$ \\
\hline
3  & $X_{1},X_{2},X_{3},X_{4},X_{6},X_{7}$  & 15  & $X_{1},X_{4},X_{7}+\alpha X_{3}$ \\
\hline
\hline
\multicolumn{2}{|c||}{r=5} & \multicolumn{1}{c|}{16 } & $X_{1},X_{2},X_{7}$ \\
\hline
1  & $X_{1},X_{2},X_{5},X_{6},X_{7}$  & 17  & $X_{1},X_{2},X_{7}+\alpha X_{3}$ \\
\hline
2  & $X_{1},X_{2},X_{3},X_{4},X_{5}+\alpha X_{7}$  & 18  & $X_{1},X_{2}+X_{3},X_{4}$ \\
\hline
3  & $X_{1},X_{2},X_{3},X_{4},X_{5}+X_{6}$  & 19  & $X_{1},X_{3},X_{4}$ \\
\hline
5  & $X_{1},X_{2},X_{3},X_{6},X_{7}+\alpha X_{4}$  & 20  & $X_{1},X_{2},X_{3}$ \\
\hline
6  & $X_{1},X_{2},X_{3},X_{4},X_{6}$  & \multicolumn{2}{c|}{r=2}\\
\hline
7  & $X_{1},X_{2},X_{3},X_{4},X_{7}$  & 1  & $X_{5}+\alpha X_{7},X_{6}$ \\
\hline
\hline
\multicolumn{2}{|c||}{r=4} & \multicolumn{1}{c|}{2 } & $X_{5},X_{7}$ \\
\hline
1  & $X_{1},X_{2},X_{5}+\alpha X_{7},X_{6}$  & 3  & $X_{6},X_{7}$ \\
\hline
2  & $X_{3},X_{4},X_{5},X_{7}$  & 4  & $X_{1},X_{6}$ \\
\hline
3  & $X_{1},X_{2},X_{5},X_{7}$  & 5  & $X_{1},X_{3}+X_{6}$ \\
\hline
4  & $X_{1},X_{3},X_{6},X_{7}$  & 6  & $X_{2},X_{3}+X_{6}+\alpha X_{4}$ \\
\hline
5  & $X_{1},X_{2},X_{6},X_{3}+\alpha X_{7}$  & 7  & $X_{3},X_{7}+\alpha X_{4}$ \\
\hline
6  & $X_{1},X_{2},X_{6},X_{7}$  & 8  & $X_{1},X_{7}+\alpha X_{3}+\beta X_{4}$ \\
\hline
7  & $X_{1},X_{2},X_{4},X_{3}+X_{6}$  & 9  & $X_{2}+X_{3},X_{4}+\alpha X_{1}$ \\
\hline
8  & $X_{1},X_{3},X_{4},X_{7}$  & 11  & $X_{3},X_{4}$ \\
\hline
9  & $X_{1},X_{2},X_{4},X_{7}+\beta X_{3}$  & 12  & $X_{1},X_{2}+X_{3}$ \\
\hline
10  & $X_{1},X_{2},X_{3},X_{4}$  & 13  & $X_{1},X_{3}$ \\
\hline
\multicolumn{2}{|c||}{r=3} & 14  & $X_{2}+\alpha X_{1},X_{3}$ \\
\hline
1  & $X_{5},X_{6},X_{7}$  & 15  & $X_{1},X_{2}$ \\
\hline
2  & $X_{3},X_{4},X_{5}+\alpha X_{7}$  & \multicolumn{2}{c|}{r=1}\\
\hline
3  & $X_{1},X_{2},X_{5}+\alpha X_{7}$  & 1  & $X_{5}+\alpha X_{7}$ \\
\hline
4  & $X_{1},X_{2},X_{5}+X_{6}$  & 2  & $X_{5}+X_{6}$ \\
\hline
5  & $X_{2}+X_{3},X_{1}-X_{4},X_{5}$  & 4  & $X_{3}+X_{6}$ \\
\hline
6  & $X_{1},X_{2},X_{5}$  & 5  & $X_{6}$ \\
\hline
7  & $X_{1},X_{6},X_{7}+\beta X_{3}$  & 6  & $X_{7}+\alpha X_{3}$ \\
\hline
8  & $X_{2},X_{3}+X_{6},X_{4}+\alpha X_{1}$  & 8  & $X_{3}$ \\
\hline
9  & $X_{1},X_{2}+X_{3},X_{6}$  & 9  & $X_{2}+X_{3}$ \\
\hline
10  & $X_{1},X_{3},X_{6}$  & 10  & $X_{1}$ \\
\hline
\end{tabular}\caption{Optimal system of subalgebras of the Lie algebra $L_{7}$.}
\label{Optimal_system}
\end{table}

\subsection{Solutions of classifining equations (\ref{eq:dec06.21.1})}

As mentioned above, substituting the constants $x_{i},\,\,\,(i=1,2,...,7)$,
determined by the basis generators of a subalgebra of the optimal
system of subalgebras of the Lie algebra $L_{7},$into classifying
equations (\ref{eq:dec06.21.1}), one obtains an overdetermined system
of equations for the function $H(x,y,t)$. Solving the overdetermined
system of equations, one finds the function $H(x,y,t)$ such that
equations (\ref{eq:SGN}) admit the corresponding Lie algebra. Considering
all subalgebras of the optimal system of subalgebras, one obtains
the group classification presented in Table \ref{Group_classification},
where representation of the function $H(x,y,t)$ is given in the second
column, and the corresponding admitted Lie algebra is presented in
the third column. The constants $k$, $l$, $\alpha$ and $\beta$
are arbitrary, the function
\[
H=lt^{2}+Q,
\]
where $l$ is an arbitrary constant and  $Q$ is an arbitrary function of its arguments. If $Q^{\prime\prime}=0$, then normally the model has more admitted generators: it is a particular case of the previous model.

Here are two typical examples.

\subsubsection{Example 1}

Consider the subalgebra 2.9: $\{X_{2}+X_{3},X_{4}+\alpha X_{1}\}$.
These generators gives the two sets of nonzero constants: for $X_{2}+X_{3}$
they are $x_{2}=1$ and $x_{3}=1$; and for $X_{4}+\alpha X_{1}$
they are $x_{1}=\alpha$ and $x_{4}=1$. The classifying equations
in this case become
\[
tH_{x}+H_{y}=a_{1}t+b_{1}\,\,\,\alpha H_{x}+tH_{y}=a_{2}t+b_{2},
\]
where $a_{i}$ and $b_{i}$ ($i=1,2$) are arbitrary constants. The
general solution of these equations is
\begin{equation}
H=y\left(\frac{-tl_{2}+l_{1}\alpha}{t^{2}-\alpha}+k_{2}\right)+x\left(\frac{-tl_{1}+l_{2}}{t^{2}-\alpha}+k_{1}\right)+Q(t),\,\,\,\alpha^{2}+l_{2}^{2}\neq0,\label{eq:example1}
\end{equation}
where $l_{1}=a_{2}-b_{1}$, $l_{2}=a_{1}\alpha-b_{2}$, $k_{1}=a_{1}$,
and $k_{2}=a_{2}$. This result is presented in Table \ref{Group_classification}
at number 15.

\subsubsection{Example 2}

Consider the subalgebra 2.1: $\{X_{5}+\alpha X_{7},X_{6}\}$. These
generators gives the two sets of nonzero constants: for $X_{5}+\alpha X_{7}$
they are $x_{5}=1$ and $x_{7}=\alpha$; and for $X_{6}$ it is $x_{6}=1$.
The classifying equations in this case become
\begin{equation}
\begin{array}{c}
(2\alpha x-y)H_{x}+(2\alpha y+x)H_{y}+\alpha tH_{t}=2\alpha H+a_{1}t+b_{1},\\
H_{t}=a_{2}t+b_{2}.
\end{array}\label{eq:dec08.21.1}
\end{equation}
Solving the second equation of (\ref{eq:dec08.21.1}), one finds $H=a_{2}\frac{t^{2}}{2}+b_{2}t+G(x,y)$,
where because of equivalence transformations one can assume that $b_{2}=0$.
Substituting the found $H$ into the first equation of (\ref{eq:dec08.21.1}),
and differentiating it with respect to $t$, one derives that $a_{1}=0$.
The first equation of (\ref{eq:dec08.21.1}) becomes
\begin{equation}
\begin{array}{c}
(2\alpha x-y)G_{x}+(2\alpha y+x)G_{y}=2\alpha G+b_{1}.\end{array}\label{eq:dec08.21.2}
\end{equation}
In polar coordinate system
\[
x=r\cos\varphi,\,\,\,y=r\sin\varphi,
\]
equation (\ref{eq:dec08.21.2}) has more convenient representation
\[
Q_{\varphi}+2\alpha rQ_{r}=2\alpha Q+b_{1}.
\]
 The general solution of the latter equation depends on $\alpha$:
\[
\alpha\neq0:\,\,\,\,G(r,\varphi)=e^{2\alpha\varphi}Q(re^{-2\alpha\varphi})-\frac{b_{1}}{2\alpha},
\]

\[
\alpha=0:\,\,\,\,G(r,\varphi)=b_{1}\varphi+Q(r),
\]
where the function $Q$ is an arbitrary function of its arguments.
These results are presented in Table \ref{Group_classification} at
number 11 and 12.

\begin{table}[h!]
\centering %
\begin{tabular}{|c|c|c|c|}
\hline
N  & $Q$  & Basis  & N(opt.syst.)\\
\hline
\hline
1  & $0$  & $X_{1},X_{2},X_{3},X_{4},X_{5},X_{6},X_{7}$  & 7.1 \\
\hline
\hline
2  & $yk_{2}+xk_{1}$, $k_{2}^{2}+k_{1}^{2}\neq0$  & $X_{1},X_{2},X_{3},X_{4},X_{6},X_{7}$  & 6.3 \\
\hline
\hline
3  & $Q(t)$, $Q^{\prime\prime\prime}\neq0$  & $X_{1},X_{2},X_{3},X_{4},X_{5}$  & 5.2$_{\alpha=0}$ \\
\hline
4  & $yk_{2}+xk_{1}+Q(t)$, $k_{2}^{2}+k_{1}^{2}\neq0$  & $X_{1},X_{2},X_{3},X_{4}$  & 4.10 \\
\hline
\hline
5  & $kr$, $k\neq0$  & $X_{5},X_{6},X_{7}$  & 3.1 \\
\hline
6  & $yk_{2}+Q(x-t^{2}/2)$  & $X_{2},X_{4},X_{6}+X_{3}$  & 3.8$_{\alpha=0}$ \\
\hline
7  & $xk_{1}+Q(y)$  & $X_{1},X_{3},X_{6}$  & 3.10 \\
\hline
8  & $xk_{1}+t^{2}Q{\displaystyle \left(\frac{y}{t^{2}}+\frac{\beta}{t}\right)}$  & $X_{1},X_{3},X_{7}+\beta X_{4}$  & 3.14$_{\alpha=0}$ \\
\hline
9  & $y{\displaystyle \left(k_{2}+\frac{l_{2}}{t}\right)}+xk_{1}+Q(t)$,
$l_{2}\neq0$  & $X_{1},X_{3},X_{4}$  & 3.19 \\
\hline
10  & $y(k_{2}t+l_{2})+xk_{1}+Q(t)$, $l_{2}\neq0$  & $X_{1},X_{2},X_{3}$  & 3.20 \\
\hline
\hline
11  & $e^{2\alpha\varphi}Q(re^{-2\alpha\varphi})$ , $Q^{\prime\prime}\neq0$  & $X_{5}+\alpha X_{7},X_{6}$, $\alpha\neq0$  & 2.1$_{\alpha\neq0}$ \\
\hline
12  & $k\varphi+Q(r)$ , $Q^{\prime\prime}\neq0$  & $X_{5},X_{6}$  & 2.1$_{\alpha=0}$\\
\hline
13  & $t^{2}Q(rt^{-2})$ , $Q^{\prime\prime}\neq0$  & $X_{5},X_{7}$  & 2.2 \\
\hline
14  & $xQ(yx^{-1})$ , $Q^{\prime\prime}\neq0$  & $X_{6},X_{7}$  & 2.3 \\
\hline
15  & (\ref{eq:example1}), $\alpha^{2}+l_{2}^{2}\neq0$  & $X_{2}+X_{3},X_{4}+\alpha X_{1}$  & 2.9 \\
\hline
16  & $y{\displaystyle \left(k_{2}+\frac{l_{2}}{t}\right)+x\left(k_{1}+\frac{l_{1}}{t}\right)}+Q(t)$,
$l_{1}\neq0$  & $X_{3},X_{4}$  & 2.11 \\
\hline
17  & $y(t^{2}k+tl_{2}+l_{0})+x(-tk+l_{1})+Q(t)$, $k\neq0$  & $X_{1},X_{2}+X_{3}$  & 2.12 \\
\hline
18  & $xk_{1}+Q(y,t)$  & $X_{1},X_{3}$  & 2.13 \\
\hline
19  & $y(tk_{2}+l_{2})+x(tk_{1}+l_{1})+Q(t)$, $l_{1}\neq0$  & $X_{1},X_{2}$  & 2.15 \\
\hline
\hline
20  & $t^{2}Q(\varphi-\ln(t)/\alpha,r/t^{2})$  & $X_{5}+\alpha X_{7}$  & 1.1$_{\alpha\neq0}$ \\
\hline
21  & $Q(\varphi-t,r)$  & $X_{5}+X_{6}$  & 1.2 \\
\hline
22  & $\varphi(l_{1}t+l_{0})+Q(t,r)$  & $X_{5}$  & 1.1$_{\alpha=0}$\\
\hline
23  & $Q(x-t^{2}/2,y)$  & $X_{3}+X_{6}$  & 1.4 \\
\hline
24  & $Q(x,y)$  & $X_{6}$  & 1.5 \\
\hline
25  & $t^{2}Q(x/t^{2},y/t^{2})$  & $X_{7}+\alpha X_{3}$  & 1.6 \\
\hline
26  & $x(k_{1}/t+k_{0})+Q(t,y)$  & $X_{3}$  & 1.8 \\
\hline
27  & $k_{1}xt+k_{2}y+Q(t,x-ty)$  & $X_{2}+X_{3}$  & 1.9 \\
\hline
28  & $x(tk_{2}+k_{1})+Q(y,t)$  & $X_{1}$  & 1.10 \\
\hline
\end{tabular}

\caption{Group classification of equations (ref{eq:SGN})}
\label{Group_classification}
\end{table}

\section{Conclusions}

The two-dimensional Green-Naghdi equations with uneven bottom topography dependent on time are studied in this paper. Group classification of these
equations with respect to the function describing the topography of
the bottom is performed in the paper. For the group classification we applied the algebraic approach.
This approach simplifies the method for solving classifying equations (\ref{eq:dec06.21.1}). For its
application one notes that the generators admitted
by equations (\ref{eq:SGN}) compose a Lie algebra which is a subalgebra
of the Lie algebra $L_{7}=\left\{ X_{1},X_{2},...,X_{7}\right\} $.
As the actions of the equivalence transformations corresponding to (\ref{eq:equiv_gen}) coincides with
the actions of the inner automorphisms of the Lie algebra $L_7$ given in Table~\ref{tab:inner}, for the
group classification one can use the optimal system of subalgebras of the Lie algebra $L_7$. Using the
generators of a chosen subalgebra from Table~\ref{Optimal_system}, one determines the coefficients
$x_i$, ($i=1,2,\ldots,7$), and substituting them into the classifying equations (\ref{eq:dec06.21.1}),
one obtains an overdetermined system of equations for the function $H(x,y,t)$. The general solution
of the latter system provides the bottom topography such that system (\ref{eq:SGN}) admits the chosen
subalgebra. The final result of the group classification is presented in Table~\ref{Group_classification}.

\section*{Acknowledgments}

The research was supported by the Russian Science Foundation Grant
No. 18-11-00238 `Hydrodynamics-type equations: symmetries, conservation laws,
invariant difference schemes'.


\begin{thebibliography}{10}

\bibitem{bk:Whitham[1974]}
G.~B. Whitham.
\newblock {\em Linear and Nonlinear Waves}.
\newblock Wiley, New York, 1974.

\bibitem{bk:Pedlosky}
J.~Pedlosky.
\newblock {\em Geophysical Fluid Dynamics}.
\newblock Springer, New York, 1987.
\newblock 2nd Edition.

\bibitem{bk:Ovsiannikov_C}
L.~V. Ovsiannikov, N.~I. Makarenko, V.~I. Nalimov, V.~Yu. Liapidevskii, P.~I.
  Plotnikov, I.~V. Sturova, V.~I. Bukreev, and V.~A. Vladimirov.
\newblock {\em Nonlinear Problems of the Theory of Surface and Internal Waves}.
\newblock Nauka, Novosibirsk, 1985.
\newblock In Russian.

\bibitem{bk:Salmon[1998]}
R.~Salmon.
\newblock {\em Lectures on Geophysical Fluid Dynamics}.
\newblock Oxford University Press, New York, 1998.

\bibitem{bk:Vallis2006}
G.~K. Vallis.
\newblock {\em Atmospheric and oceanic fluid dynamics. Fundamentals and
  large-scale circulation}.
\newblock Cambridge University Press, Cambridge, 2006.

\bibitem{bk:PetrosyanBook2010}
A.~S. Petrosyan.
\newblock {\em Additional chapters of heavy fluid hydrodynamics with a free
  boundary}.
\newblock Space Research Institute of the Russian Academy of Sciences, Moscow,
  2014.
\newblock in Russian.

\bibitem{bk:BonnetoBarthelemy_et[2011]}
P.~Bonneton, E.~Barth\'elemy, F.~Chazel, R.~Cienfuegos, D.~Lannes, F.~Marche,
  and M.~Tissier.
\newblock Recent advances in {S}erre-{G}reen-{N}aghdi modelling for wave
  transformation, breaking and runup processes.
\newblock {\em Euro. J. Mech. B/Fluids}, 30:589--597, 2011.

\bibitem{bk:KhakimzyanovDutykhFedotovaMitsotakis}
G.~S. Khakimzyanov, D.~Dutykh, Z.~I. Fedotova, and D.~E. Mitsotakis.
\newblock Dispersive shallow water wave modelling. {P}art {I}: {M}odel
  derivation on a globally flat space.
\newblock {\em Commun. Comput. Phys.}, 23(1):1--29, 2018.

\bibitem{bk:Matsuno[2015]}
Y.~Matsuno.
\newblock Hamiltonian formulation of the extended {G}reen-{N}aghdi equations.
\newblock {\em Phycs D}, 301-302:1--7, 2015.

\bibitem{bk:Serre[1953]}
F.~Serre.
\newblock Contribution \`a l'\'etude des \'ecoulements permanents et variables
  dans les canaux.
\newblock {\em Houille Blanche}, 3:374--388, 1953.

\bibitem{bk:SuGardner[1969]}
C.~H. Su and C.~S. Gardner.
\newblock {K}orteweg-de {V}ries equation and generalizations. {III}.
  {D}erivation of the {K}orteweg-de {V}ries equation and {B}urgers equation.
\newblock {\em Journal of Mathematical Physics}, 10(3):10--23, 1969.

\bibitem{bk:KaptsovMeleshkoSamatova2021}
E.~I. Kaptsov, S.~V. Meleshko, and N.~F. Samatova.
\newblock The one-dimensional green-naghdi equations with a time dependent
  bottom topography and their conservation laws.
\newblock {\em Physics of Fluids}, 32(12):123607, 2020.

\bibitem{bk:Hammack1973}
J.~L. Hammack.
\newblock A note on tsunamis: their generation and propagation in an ocean of
  uniform depth.
\newblock {\em J . Fluid Mech.}, 60:769--799, 1973.
\newblock part 4.

\bibitem{bk:DutykhDias2007b}
D.~Dutykh and F.~Dias.
\newblock Water waves generated by a moving bottom.
\newblock In A.~Kundu, editor, {\em Tsunami and Nonlinear Waves}, pages 65--96.
  Springer-Verlag (Geo Sc.), Berlin, Heidelberg, 2007.

\bibitem{bk:Ovsiannikov1978}
L.~V. Ovsiannikov.
\newblock {\em Group Analysis of Differential Equations}.
\newblock Nauka, Moscow, 1978.
\newblock {E}nglish translation, {A}mes, {W}.{F}., Ed., published by Academic
  Press, New York, 1982.

\bibitem{bk:Olver[1986]}
P.~J. Olver.
\newblock {\em Applications of {L}ie Groups to Differential Equations}.
\newblock Springer-Verlag, New York, 1986.

\bibitem{bk:Ibragimov[1983]}
N.~H. Ibragimov.
\newblock {\em Transformation Groups Applied to Mathematical Physics}.
\newblock Nauka, Moscow, 1983.
\newblock {E}nglish translation, Reidel, D., Ed., Dordrecht, 1985.

\bibitem{bk:BlumanKumei1989}
G.~W. Bluman and S.~Kumei.
\newblock {\em Symmetries and Differential Equations}.
\newblock Springer-Verlag, New York, 1989.

\bibitem{bk:HandbookLie}
N.~H. Ibragimov, editor.
\newblock {\em {CRC} Handbook of {L}ie Group Analysis of Differential
  Equations}, volume 1, 2, 3.
\newblock CRC Press, Boca Raton, 1994, 1995, 1996.

\bibitem{bk:Gaeta1994}
G.~Gaeta.
\newblock {\em Nonlinear Symmetries and Nonlinear Equations}.
\newblock Kluwer, Dordrecht, 1994.

\bibitem{bk:BagderinaChupakhin[2005]}
Yu.~Yu. Bagderina and A.~P. Chupakhin.
\newblock Invariant and partially invariant solutions of the {G}reen-{N}aghdi
  equations.
\newblock {\em Journal of Applied Mechanics and Technical Physics},
  46(6):791--799, 2005.

\bibitem{bk:SiriwatKaewmaneeMeleshko2016}
P.~Siriwat, C.~Kaewmanee, and S.~V. Meleshko.
\newblock Symmetries of the hyperbolic shallow water equations and the
  {G}reen-{N}aghdi model in {L}agrangian coordinates.
\newblock {\em International Journal of Non-Linear Mechanics}, 86:185--195,
  2016.

\bibitem{bk:DorodnitsynKaptsovMeleshko2021}
V.~A. Dorodnitsyn, E.~I. Kaptsov, and S.~V. Meleshko.
\newblock Symmetries and difference schemes of the 1{D} {G}reen-{N}aghdi
  equations.
\newblock {\em Journal of Nonlinear Mathematical Physics}.
\newblock in press.

\bibitem{bk:GrigorievMeleshkoSuriyawichitseranee2012}
Yu. N.Grigoriev, S.~V. Meleshko, and A.~Suriyawichitseranee.
\newblock On the equation for the power moment generating function of the
  {B}oltzmann equation. group classification with respect to a source function.
\newblock In O.O. Vaneeva, C.~Sophocleous, R.O. Popovych, P.G.L. Leach, V.M.
  Boyko, and P.A. Damianou, editors, {\em Group Analysis of Differential
  Equations \& Integrable Systems}, pages 98--110. University of Cyprus,
  Nicosia, 2012.

\bibitem{bk:Chirkunov[2012]}
Yu.~A. Chirkunov.
\newblock Generalized equivalence transformations and group classification of
  systems of differential equations.
\newblock {\em Journal of Applied Mechanics and Technical Physics},
  53(2):147--155, 2012.

\bibitem{bk:Kasatkin[2012]}
A.~A. Kasatkin.
\newblock Symmetry properties for systems of two ordinary fractional
  differential equations.
\newblock {\em Ufa Mathematical Journal}, 4(1):71--81, 2012.

\bibitem{bk:MkhizeMoyoMeleshko[2014]}
T.~G. Mkhize, S.~Moyo, and S.~V. Meleshko.
\newblock Complete group classification of systems of two linear second-order
  ordinary differential equations. {A}lgebraic approach.
\newblock {\em Mathematical Methods in the Applied Sciences}, 38:1824--1837,
  2015.

\bibitem{bk:OpanasenkoBoykoPopovych_2020}
S.~Opanasenko, V.~Boyko, and R.~O. Popovych.
\newblock Enhanced group classification of nonlinear diffusion-reaction
  equations with gradient-dependent diffusivity.
\newblock {\em Journal of Mathematical Analysis and Applications},
  484(1):123739, 2020.

\bibitem{Bazdenkov_C_1987}
S.~V. Bazdenkov, N.~N. Morozov, and O.~P. Pogutse.
\newblock Dispersive effects in two-dimensional hydrodynamics.
\newblock {\em Sov. Phys. Dokl.}, 32:262--264, 1987.

\bibitem{bk:LiapidevskiiGavrilova2008}
V.~Yu. Liapidevskii and K.~N. Gavrilova.
\newblock Dispersion and blockage effects in the flow over a sill.
\newblock {\em J Appl Mech Tech Phys}, 49(7):34--45, 2008.

\bibitem{bk:Matsuno[2016]}
Y.~Matsuno.
\newblock Hamiltonian structure for two-dimensional extended {G}reen-{N}aghdi
  equations.
\newblock {\em Proceedings of the Royal Society. Mathematical, physical and
  engineering sciences}, 472(2190), 2016.

\bibitem{bk:Meleshko[2005]}
S.~V. Meleshko.
\newblock {\em Methods for Constructing Exact Solutions of Partial Differential
  Equations}.
\newblock Mathematical and Analytical Techniques with Applications to
  Engineering. Springer, New York, 2005.

\bibitem{bk:Hearn}
A.~C. Hearn.
\newblock {\em REDUCE Users Manual, ver. 3.3}.
\newblock The Rand Corporation CP 78, Santa Monica, 1987.

\bibitem{bk:Ovsiannikov[1993opt]}
L.~V. Ovsiannikov.
\newblock On optimal system of subalgebras.
\newblock {\em Dokl. RAS}, 333(6):702--704, 1993.

\bibitem{bk:Ovsiannikov[1994]}
L.~V. Ovsiannikov.
\newblock Program {SUBMODELS}. {G}as dynamics.
\newblock {\em J. Appl. Maths Mechs}, 58(4):30--55, 1994.

\end{thebibliography}

\end{document}